\documentstyle[12pt]{article}

\begin{document}

\topmargin 0pt
\oddsidemargin 5mm
\def\bbox{{\,\lower0.9pt\vbox{\hrule \hbox{\vrule height 0.2 cm
\hskip 0.2 cm \vrule height 0.2 cm}\hrule}\,}}

\newcommand{\hs}[1]{\hspace{#1 mm}}
\newcommand{\shalf}{\frac{1}{2}}
\newcommand{\pa}{\partial}
\newcommand{\tri}{{\small $\triangle$}}
\newcommand{\dz}{\frac{dz}{2\pi i}}
\newcommand{\ra}{\rightarrow}
\newcommand{\la}{\leftarrow}
\newcommand{\nn}{\nonumber \\}
\def\a{\alpha}
\def\b{\beta}
\def\g{\gamma}
\def\G{\Gamma}
\def\d{\delta}
\def\D{\Delta}
\def\e{\epsilon}
\def\ve{\varepsilon}
\def\z{\zeta}
\def\t{\theta}
\def\vt{\vartheta}
\def\r{\rho}
\def\vr{\varrho}
\def\k{\kappa}
\def\l{\lambda}
\def\L{\Lambda}
\def\m{\mu}
\def\n{\nu}
\def\o{\omega}
\def\O{\Omega}
\def\s{\sigma}
\def\vs{\varsigma}
\def\S{\Sigma}
\def\vphi{\varphi}
\def\av#1{\langle#1\rangle}
\def\na{\nabla}
\def\hg{\hat g}
\def\un{\underline}
\def\ov{\overline}
\def\tr{{\rm tr}}
\def\sabs{\s_{\rm abs}}
\def\w{\wedge}

\newcommand{\ap}[1]{Ann.\ Phys.\ {\bf #1}}
\newcommand{\np}[1]{Nucl.\ Phys.\ {\bf  B#1}}
\newcommand{\pl}[1]{Phys.\ Lett.\ {\bf  B#1}}
\newcommand{\cmp}[1]{Comm.\ Math.\ Phys.\ {\bf #1}}
\newcommand{\pr}[1]{Phys.\ Rev.\ {\bf  D#1}}
\newcommand{\prl}[1]{Phys.\ Rev.\ Lett.\ {\bf #1}}
\newcommand{\prp}[1]{Prog.\ Theor.\ Phys.\ {\bf #1}}
\newcommand{\ptps}[1]{Prog.\ Theor.\ Phys.\ Suppl.\ {\bf #1}}
\newcommand{\mpl}[1]{Mod.\ Phys.\ Lett.\ {\bf #1}}
\newcommand{\ijmp}[1]{Int.\ Jour.\ Mod.\ Phys.\ {\bf #1}}
\newcommand{\cqg}[1]{Class.\ Quant.\ Grav.\  {\bf #1}}
\newcommand{\prep}[1]{Phys.\ Rep.\ {\bf #1}}
\newcommand{\rmp}[1]{Rev.\ Mod.\ Phys.{\bf #1}}

\begin{titlepage}
\setcounter{page}{0}

\begin{flushright}
COLO-HEP-385 \\
hep-th/9706142 \\
June 1997
\end{flushright}

\vspace{5 mm}
\begin{center}
{\large A note on brane creation }
\vspace{10 mm}

{\large S. P. de Alwis\footnote{e-mail:  
dealwis@gopika.colorado.edu}}\\
{\em Department of Physics, Box 390,
University of Colorado, Boulder, CO 80309}\\
\vspace{5 mm}
\end{center}
\vspace{10 mm}

\centerline{{\bf{Abstract}}}
The M-theory origin of brane creation processes is discussed.

\end{titlepage}
\newpage
\renewcommand{\thefootnote}{\arabic{footnote}}
\setcounter{footnote}{0}

\setcounter{equation}{0}
Recently Hanany and Witten\cite{hw} have shown how to derive  
non-perturbative
results for (three-dimensional) field theory starting from  
ten-dimensional type
IIB. In the course of the discussion they  derived certain brane  
transition
rules which show that when certain branes cross,  other  branes are  
created.
This phenomenon has been the subject of some discussion
\cite{aehw},\cite{bdg},\cite{dfk},\cite{bgl},\cite{hz}. In this paper  
we will
discuss the  M-theory origin of these transition rules. In particular   
the
correct normalizations of various terms in the action of  five and  
two branes
coupled to M-theory obtained in \cite{ew}, \cite{sda},  will enable  
us to fix
directly the relations between the charges and intersection numbers  
that enter
into these
rules.

 We work in eleven dimesional Planck units, i.e. we have set  
$2\k_{11}^2=1$ so
that the M-membrane tension is $T_2=(2\pi)^{2/3}$ and the five-brane  
tension is
$T_5= (2\pi )^{1/3}$.\footnote{These results follow from the Dirac  
quantization
condition and the relation $T_2^2/T_5=2\pi$ and are reviewed in  
\cite{sda}.}
The action $I$ for low-energy M theory coupled to a two-brane and a  
five-brane
allowing also for the
possibility of the two-brane ending  on the five-brane  
is\footnote{The complete
form of this action as written here is given in \cite{sda}. It  
depends heavily
on the work of other authors, particularly \cite{ew}. A complete (to  
the
authors knowledge) list of references is given in \cite{sda}.  },
\begin{eqnarray}\label{action}
{I\over 2\pi}&=&-\shalf (2\pi)^{-{1\over 3}}\int_{M}x_4\w  
*x_4+{1\over
3}\int_Mc_3\w x_4'\w
 x_4'-\int_{M}x_4\w (\shalf x_4'\w c_3-\O_7+\t_7)\nn &
&-\int_{W_3}c_3+\int_{\pa W_3} b_2 -{1\over 4}\int_{W_6}h\w
*h-\int_{W_6}\shalf x_4'\w b_2,
\end{eqnarray}
In the above $M$ is the eleven-manifold, $W_3$ is the (open) world  
volume of
the two-brane whose boundary sits on $W_6$ (the world volume of the
five-brane). Also\footnote{The terms $\O_7$ and $\t_7$ are related to
perturbative anomaly cancellation coming from five-brane anomalies.  
They are
irrelevent for us and hence will be ignored in the rest of this  
paper.}
$x_4=x_4'+\t_4$, where the second term is a coexact four-form that  
solves
$dx_4=d\t_4 =-\d_5(M\rightarrow W_6)$. In the neighbourhood of the  
five-brane
the $M$ has  the (twisted) product form
$W_6\otimes D_5$ where $D_5$ is an open five disc bounded by a four  
sphere and
the delta function is normalized such that $\int_D\d  
=\int_{S_4}x_4=1$.  It
should be noted that the fields in the above action are normalized  
such that
$\int x'_4$ over any four cycle is integral. Also $h_3 = h_3'  
-c_3...$ where
$h_3'=db_2$ locally, and  the normalization is such that   $\int  
h_3'$ over any
three cycle in $W_6$ is integral. The ellipses in the expression for  
$h_3$ come
from the presence of sources on the five-brane due to the ends of  
two-branes.
This is fixed by comparing with the equation of motion coming from  
the above
action and imposing the self-duality constraint $*h=h$.

The important point about this action is that it couples both the two  
brane,
and its magnetic dual
the five-brane, to the M-theory background given in its normal form  
(i.e. with
just the three form gauge field). It also contains the terms relevant  
to having
2-branes with boundaries on 5-branes. In this sense there is no  
analogue  in
string theory effective actions or four dimensional field theory of  
charges and
monopoles.
To derive the equations of motion it is convenient to rewrite the  
last four
integrals as follows.
\begin{eqnarray}\label{}
&-&\int_Mc_3\w \t_8(M\ra W_3) + \int_M b_2\w\d_5 (M\ra  
W_6)\w\d_4(W_6\ra \pa
W_3)\nn
&-&{1\over 4}\int _M(h\w *h)\w \d_5(M\ra W_6)+\int_M\shalf x'_4\w  
b_2\w\d_5
(M\ra W_6)
\end{eqnarray}
In the above the $\d_r$ denotes delta function r-forms with the  
indicated
support, and $\t_r$ is an r-form which is a  product of theta  
functions and
delta funtions which restrict $M$ to the open manifold $W_3$. This  
function
satisfies $d\t_8 (M\ra W_3) =-\d_9(M\ra \pa W_3)$ (see  for example  
section 6
of \cite{sda}).
The variation with respect to $b_2$ then gives,
\begin{equation}\label{h}
d*h =dh=-x_4'|_{W_6} - 2\d_4(W_6\ra\pa W_3)
\end{equation}
where in the first equality we used the self duality of $h$.
It is also instructive to derive the $c_3$ equation of motion.
\begin{equation}\label{}
(2\pi)^{-{1\over 3}}d*x_4 = -\shalf x_4'\w x_4'-\t_8(M\ra W_3)-\t_4\w
x_4'+\shalf h\w\d_5+\shalf (db_2-c_3)\w\d_5
\end{equation}
It may be checked that this satisfies the consistency condition  
$d^2*x_4=0$
once one uses the
Bianchi identity for $h$ (\ref{h}) and the equation $d\t_4=-\d_5$.

Let us now consider the case of several three-branes ending on a  
given
five-brane. In this case the second term on the right hand side of  
(\ref{h})
has to be replaced by a sum over all the string sources on the  
five-brane,
coming from the boundaries of the membranes. Thus we get
\begin{equation}\label{}
dh=-x_4'|_{W_6} - 2\sum_i e_i\d_4(W_6\ra\pa W^i_3)
\end{equation}
where $e_i=\pm 1$ are the (normalized) charges carried by these  
sources.
Integrating over
any four cycle $Y_4$ in $W_6$ we get,
\begin{equation}\label{Y}
\int_{Y_4\subset W_6} x_4'=2\sum_{i\subset Y_4}e_i
\end{equation}
Note that the last sum is only over those end strings that are  
contained within
$Y_4$.
Now $x_4'$ is the ambient field at $W_6$ which may have as sources  
all other
five-branes apart from the one that is being considered. In  
particular the
above integral  is the intersection number of all five-branes  
threading the
four cycle $Y_4$. For instance if a five brane with world volume  
$W^i_6$
threaded
the four cycle then,
\begin{equation}\label{}
\int_{Y_4}x_4'=\pm\int_{D_5}dx_4'=\pm\int_{D_5}\d (M\ra W^i_6),
\end{equation}
 where $D_5\subset M$ is an open disc with boundary $\pa D = Y_4$  
except for
orientation which may be the same or opposite. Thus the
equation (\ref{Y}) can be written as,
\begin{equation}\label{Q}
Q=\shalf\sum_{i\subset D_5}e_i^{(5)}-\sum_{i\subset Y_4}e_i^{(2)}=0
\end{equation}
where $e_i^{(5)}=\pm 1$ are the five-brane charges and the first sum  
extends
over all five branes intersecting $D_5$.

The above considerations are valid for the compact case. In the  
non-compact
case which is considered in the literature (\cite{hw})  the integral  
$\int dh$
is not zero but is an integral over the surface at infinity and so  
$Q$ is a
constant rather than zero, as in the analogous case considered in  
\cite{hw}.
The important point about the formula (\ref{Q}) is that  the correct  
relative
normalizations of the different terms which is  a consequence of   
gauge
invariance and the tension formulae, guarantee that the five-brane
intersections contribute half as much as the
end strings  which bound the membranes on the five-brane.

The  M-theory relation (\ref{Q}) implies similar relations for   
various string
theory configurations by various S- and T-duality transformations.  
For instance
let us  consider the M theory configuration of two five-branes with a  
two-brane
suspended between them (i.e. with each of  two boundaries sitting  on  
one of
the five-branes). The coordinates in parantheses are the spatial  
directions in
which the branes are extended. (The M-coordinates are
$(x^0,x^1,\ldots,x^{10})$.)
\begin{equation}
M:~~~5M~(x^1,x^3,x^4,x^5, x^{10});~~~2M~(x^1,  
x^6),~~~5M~(x^1,x^2,x^7,x^8,x^9)
\end{equation}
By wrapping $x^{10}$ around a cicle and letting its radius shrink to  
zero we
have the corresponding IIA configuration of a two(D)-brane suspended  
between a
four(D)-brane and a five(NS)-brane.
\begin{equation}
IIA:~~~4D~(x^1,x^3,x^4,x^5);~~~2D  
(x^1,x^6);~~~5NS~(x^1,x^2,x^7,x^8,x^9)
\end{equation}
By T-dualizing along $x^2$ we then get the following IIB  
configuration.
\begin{equation}
IIB:~~~5D~(x^1,x^2,x^3,x^4,x^5);~~~2D
(x^1,x^2,x^6);~~~5NS~(x^1,x^2,x^7,x^8,x^9)
\end{equation}
This is the (S-dual of the) configuration considered by  Hanany and  
Witten.
Successive T- and S- dualities (U-duality) will enable us to generate  
many
different configurations considered in the literature.  For instance  
as pointed
out in \cite{bgl} starting from the last  configuration, one can  
first
T-dualize with respect to $(x^1,x^2)$ then S dualize and then  
T-dualize with
respect to $(x^3,x^4,x^5)$ to get the  IIA configuration of a  
0D-brane and a
8D-brane
with a fundamental string suspended between them. This can also be  
obtained
directly from the
following  M-theory configuration.
\begin{equation}\label{}
M:~~~5M~(x^1,x^2,x^3,x^4,x^{10});~~~2M~(x^5,x^{10})~~5M~(x^6,x^7,x^8,x 
^9,x^{10})
\end{equation}
By compactifying $x^{10}$ on $S^1$ we get two 4D-branes with  
fundamental string
suspended between them, and then T-dualizing with respect to
$(x^6,x^7,x^8,x^9)$ we get an 8D-brane and a 0D-brane
with an F-string suspended between them. This result however seems to  
indicate
that the latter configuration which is contained in massive IIA can  
be obtained
from M-theory
which does not permit a cosmological constatnt. Presumably this  
happens because
(in both ways) of getting this configuration one has to go through a  
nine
(non-compact) dimensional theory that has to be obtained by a  
non-trivial
Scherk-Schwarz type  compactification from
10 dimensions as discussed in \cite{bdgpt}. It would be interesting  
to
investigate this issue further.

In the above we've used compactification of non-compact directions  
and
T-duality at will. However it is worthwhile pointing out that this  
cannot
always be done,
  by showing how a potential paradox is avoided. This relates to the  
fact that
total charge  of a configuration can be non-zero only if the  space  
is
non-compact. First consider  nine-branes
in type IIB string theory.  As is well-known the only possibility is  
when one
has imposed a  $Z_2$
projection to get the type I string from IIB so that the  
corresponding single
orientifold plane charge cancels the charge of 16 nine-branes and  
their images
to give an SO(32) theory(see \cite{jp} for a review). This is the  
case even if
the space is non-compact.
The point is that (for $n$ nine-branes) there will be a term
$n\int_{M_{10}}A_{10}$ without a corresponding kinetic term, since
$dA_{10}=F_{11}=0$  identically  in 10 dimensions, so that  
integration over the
10-form gauge field gives $n=0$ except in the above mentioned type I  
case.  The
 puzzle is that one seems to be able to obtain a configuration of $n$  
nine
branes starting from the corresponding configuration of lower  
dimensional
branes, in a non-compact space, and T dualizing. The resolution is   
that
T-duality  requires that the transverse dualized dimension actually  
be compact,
while the existence of an arbitrary number of p-D-branes (with $p<9$)  
requires
the existence of at least one transverse dimension that is  
non-compact. Thus
for instance consider the case of $n$ parallel 8-D-branes in type IIA
perpendicular to the $X^9$ axis. The relevant piece of the action is
\begin{equation}\label{}
\shalf\int_{M_{10}}F_{10}\w *F_{10}  
+\sum_{i=1}^n\e^i\mu_8\int_{W^i_9} A_9
\end{equation}
 where  $\e^i=\pm$ and $\mu_8$ is the magnitude of the eight-brane  
charge and
$W^i_9$ is the world volume of the i'th 8-brane. The equation of  
motion for
$A_9$ is $d*F_{10}=\mu_8\sum\e^i\d (M_{10}\rightarrow W^i_9)$.  
Integrating over
a disc transverse to the eight-branes we get
\begin{equation}\label{}
Q=\int_{S^0_{\infty}}*F_{10}=*F_{10}(X^9=+\infty  
)-*F_{10}(X^9=-\infty
)=\mu_8\sum\e^i
\end{equation}
If the ninth dimension is non-compact then clearly we can have  
non-zero (total)
8-brane charge. But in this case we cannot T-dualize in this  
direction. On the
other hand if  this dimension is compact  the total charge is zero  
and we can
T-dualize to get the IIB situation discussed earlier.
Thus there is no conflict!

 We end this note by pointing out that the above mentioned problem  
does not
occur for the
T-dualities  that were used earlier to relate various brane  
configurations.
In each case that we use T-duality there is at least 

one  dimension
transverse to each brane that remains
uncompactified, so that the charge is not forced to vanish.

\noindent{\bf Acknowledgments}\\
I wish to acknowledge  correspondence with Oren Bergman. This work is  
partially
supported by the Department of Energy contract No.  
DE-FG02-91-ER-40672.


\end{document}